# Polarization spectroscopy of defect-based single photon sources in ZnO

N. R. Jungwirth, H. -S. Chang, M. Jiang, and G. D. Fuchs

Cornell University, Ithaca, New York 14853, USA

**Abstract:**

Point-defects in wide bandgap semiconductors are promising candidates for future applications that necessitate quantum light sources. Recently, defect-based single photon sources have been observed in ZnO that are very bright and remain photoactive from 4.5 K to room temperature. Despite several investigations, the structure and electronic states of these emitters remain unknown. In this work, we establish a procedure to distinguish a Z dipole from an XY dipole when studying quantum emitters that are randomly oriented. Our cryogenic and room temperature polarization measurements collectively establish that these unidentified ZnO quantum emitters have a Z dipole. We show that the associated absorption and emission dipoles are parallel within experimental uncertainty for all 32 individuals studied. Additionally, we apply group theory and find that assuming the defect symmetry belongs to a point-group relevant to the ZnO wurtzite lattice, the ground and excited states are orbital singlets. These results are a significant step in identifying the structure and electronic states of defect-based single photon sources in ZnO.

Point-defects in semiconductors have been identified as single photon sources (SPSs) with prospective applications in precision sensing and quantum communication.[1–5] While isolated defects in diamond including the nitrogen-vacancy (NV) center and the silicon-vacancy center have garnered the most attention,[5–11] recent efforts to find viable defect-based SPSs in conventional semiconductors[1] have uncovered promising candidates in SiC[12–15] and ZnO.[16–20] Unlike diamond and SiC, ZnO has the advantage of a direct bandgap, thus offering the possibility of incorporating point-defects with optoelectronic devices. Additionally, piezoelectricity in ZnO introduces the prospect of directly utilizing lattice strain to control single defects.[21,22] These enticing bulk properties of ZnO are complemented by a wealth of established growth methods that would facilitate fabrication of photonic devices that exploit single-defect properties.[23–29] Furthermore, the quantum emission from ZnO point-defects can be very bright (>100 kPhotons/s) with high polarization visibility,[18] which is advantageous for applications requiring high bandwidth and/or polarized single photons on demand.

Despite these attractive properties, the progress in realizing quantum photonic devices based on this platform is hampered by the absence of a detailed understanding of the defect's structure and electronic states. In our previous study, we found significant defect-to-defect variability in the excited state lifetime ($1 - 13\ ns$), emission spectrum, fluorescence intensity, and photodynamics,[17] which makes identification of the defect a challenge. Moreover, despite a wealth of prior research into ZnO defect fluorescence,[29–32] there is yet no correspondence between the experimental observations of quantum emitters in ZnO and the theoretically predicted behavior of candidate defect structures. Establishing the identity of ZnO quantum emitters could enable creation of high-quality SPSs in a readily engineered material and provide the framework required to address the previously observed variability. A key step is to perform experiments that shed light on the properties of the defect's electronic ground and excited state.

In this work we establish a procedure for distinguishing a "Z dipole" from an "XY dipole" in isolated quantum emitters that have unknown orientation because they reside in unordered nanostructures. Then, in concert with group theoretic considerations, we determine which electronic states could be responsible for our experimental observations. Our approach involves measuring polarization properties of absorption and emission from many isolated defects with different orientations and comparing the results with the expectations for each dipole type. Our cryogenic and room temperature measurements reveal that, unlike the NV center in diamond,[33–36] the ZnO defects we study have a Z dipole as opposed to an XY dipole. Across the 32 defects investigated, we find that the associated absorption and emission dipoles are parallel within experimental uncertainty. The identification of a Z dipole is interpreted in the context of point-group theory, enabling us to infer the allowed properties of the defect's ground and excited state wavefunctions for several symmetries common to the ZnO wurtzite lattice. These results constitute a significant step towards identifying the structure and electronic states of SPSs in ZnO.

Figure 1 depicts the house-built confocal microscope used to perform polarization spectroscopy of isolated defects in randomly oriented ZnO nanoparticles and ⟨001⟩ oriented sputtered ZnO films (see Supplementary Information for sample information). For absorption measurements, a fixed polarizer (FP1) followed by a rotatable half wave plate (HWP1) was used to produce an arbitrary linear polarization state of the exciting light. The exciting light is directed to a high NA (0.7) microscope objective (MO) by a dichroic mirror (DM). Between HWP1 and the MO, the exciting light inherits some ellipticity from retardances introduced by the excitation path. Using a variable compensator, we measure the net excitation path retardance at 532 nm to be ~0.21 waves. We correct for this using a fixed wave plate (FWP1) selected to give an appropriate correcting retardance. Though all data presented here are for defects excited with 532 nm light, we performed identical measurements using 473 nm light that was likewise compensated for path retardances, and found consistent results (see Supplementary Information).

The exciting light may be scanned over the sample in search of isolated defects. A defect with symmetry axis *D* may absorb the exciting light and emit polarized single photons into the focal cone of the objective. These collected photons range from 560 - 720 nm and each inherits a wavelength dependent retardance in the collection path. At 630 nm the collection path retardance is ~0.28 waves. This is compensated by FWP2 and the polarization state may be rotated by achromatic HWP2. Each collected photon encounters a polarization insensitive 50/50 beam splitter (BS) and is directed towards an avalanche photodiode (APD) sensitive to single photons. Photons directed towards APD1 may be detected independent of their polarization whereas photons directed towards APD2 pass through FP2 with a probability determined by their polarization state. Time correlated single photon counting (TCSPC) enables measurement of $g^{(2)}(\tau)$ to verify a single defect is being probed. The polarization state of excitation (emitted) light was calibrated by measuring 532 nm (630 nm) light with a polarization analyzer placed at the objective for various orientations of HWP1 (HWP2). The polarization extinction ratio of the exciting (collected) light is greater than 1000 (10) for all positions of HWP1 (HWP2). The diminished extinction ratio for collected light is because the emission spectrum extends over a broad range (~300 meV) compared to the monochromatic exciting light.

    Our polarization measurements may be understood in terms of absorption and emission dipoles. In general, point-defects in crystals are associated with an axis of symmetry $\boldsymbol{D}$ that corresponds to a distinct crystallographic direction. In such systems there are two electric dipole transitions relevant to polarization: $\pi$ transitions and $\sigma$ transitions. For $\pi$ transitions the absorption dipole $\boldsymbol{\mu}_A$ is parallel to $\boldsymbol{D}$ and the absorption amplitude is proportional to $\cos^2 \theta_A$, where $\theta_A$ is the angle between $\boldsymbol{\mu}_A$ and the incident electric field $\boldsymbol{E}_{exc}$. Analogously, the fluorescence intensity detected through a linear polarizer is proportional to $\cos^2 \theta_E$, where $\theta_E$ is the angle between the polarizer and the emission dipole $\boldsymbol{\mu}_E$.[37] Here we define the

$z$ axis as the axis of the objective and, for a particular defect, the $x$ axis as being perpendicular to $\boldsymbol{D}$ and $z$ (see Figure 1). If the emission and absorption dipoles are parallel, then in spherical coordinates, with polar angle $\theta$ and azimuthal angle $\phi$, we have $\boldsymbol{\mu}_A = \boldsymbol{\mu}_E = \boldsymbol{D} = \sin\theta\,\hat{j} + \cos\theta\,\hat{k}$. The fluorescence $F_\pi$ detected when the exciting light is polarized at $\phi_{exc}$ and the collection polarizer is positioned at $\phi_{coll}$ is then proportional to

$$F_\pi \propto (\sin\theta \sin\phi_{exc})^2 (\sin\theta \sin\phi_{coll})^2. \tag{1}$$

Eq 1 assumes that both the collected and exciting light propagate in the z direction and thus ignores the effect of the high NA objective, which we account for below.

For $\sigma$ transitions the absorption probability is proportional to $\sin^2\theta_A$, where $\theta_A$ is the angle between $\boldsymbol{D}$ and $\boldsymbol{E}_{exc}$. Similarly, the fluorescence intensity detected through a polarizer is proportional to $\sin^2\theta_E$, where $\theta_E$ is the angle between the polarizer and $\boldsymbol{D}$.[37] Equivalently, $\sigma$ transitions may be described by two mutually perpendicular dipoles that span the plane normal to $\boldsymbol{D}$. The absorption and emission probabilities for each of these orthogonal dipoles are precisely those given above for $\pi$ transitions.[38] Because $\pi$ ($\sigma$) transitions are associated with one (two) dipole(s), such transitions are referred to as having a Z (XY) dipole.[39]

Figure 2 is a representative set of polarization measurements made on a single ZnO defect. Monitoring the count rate on APD1 while rotating HWP1 enables measurement of the absorption profile of single defects (Figure 2a). After setting HWP1 to its maximum position, monitoring the count rate on APD2 while rotating HWP2 enables us to measure the polarization state of the emitted light (Figure 2b). Figures 2a and 2b both present a maximum at ~90° and a minimum at ~0°, suggesting the absorption and emission dipoles are parallel. We also rotate HWP1 and HWP2 simultaneously such that the effective polarization of the exciting light and the light collected at APD2 are parallel (Figure 2c) or perpendicular (Figure 2d). All polarization plots can be fit using eq 1 for a Z dipole, or the XY analog, with a constant background added. However, the background value that yields the best-fit substantially exceeds the background fluorescence measured experimentally. The discrepancy in the background level predicted using eq 1 and that measured experimentally is explained by the loss of polarization visibility when imaging with a high NA objective. This is seen by inspecting eq 1, which indicates that $F_\pi = 0$ whenever $\phi_{exc}$ or $\phi_{coll}$ is 0°. We detect non-negligible fluorescence at $\phi_{exc} = 0°$ in Figure 2a and at $\phi_{coll} = 0°$ in Figure 2b. These photons that are not predicted by eq 1 come from two sources: unpolarized background fluorescence and polarized defect fluorescence whose polarization is obfuscated by the high NA objective.[40]

Following the method proposed by Fourkas,[40] we developed an analytical model for Z and XY dipoles that accounts for our high NA objective. We simultaneously fit all plots in Figure 2 using our model by introducing the 3-dimensional defect orientation and collection path extinction ratio as free parameters. The nearly negligible background fluorescence, which is difficult to precisely measure, is not included as a free parameter. The resulting best-fits assuming a Z and XY dipole are shown as the solid and dashed lines in Figure 2, respectively. The Z (XY) dipole best-fit occurs for a collection path extinction ratio of 20 ($\sim 10^8$), which is consistent (inconsistent) with previous calibration measurements. Moreover, while the Z dipole model fits all plots adequately, the XY dipole model fails to fit Figure 2d as well as the Z dipole model does.

Figure 2 makes a strong case for SPSs in ZnO having a Z dipole but does not eliminate the XY dipole possibility. To strengthen the Z dipole argument, we acquired polarization measurements for 21 defects in randomly oriented NPs, which allowed us to sample various orientations of the defect symmetry axis $\boldsymbol{D}(\theta, \phi)$. Every such measurement was qualitatively similar to those shown in Figure 2.

Here we explain why the plots in Figure 2, that are representative of all investigated defects, support a Z dipole. Because single photon counting experiments are shot noise limited, dim emitters can be difficult to identify over the background fluorescence. Therefore, the defect orientations that produce the greatest photon yield are precisely those that are most likely to be identified for study. Figure 3 displays what our Z and XY dipole models predict for the perpendicular measurement of Figure 2d for defect orientations $\{\theta, \phi\} = \{0°, 0°\}$, $\{30°, 0°\}$, $\{60°, 0°\}$, and $\{90°, 0°\}$. Each curve assumes the same intrinsic oscillator strength and is therefore proportional to the fluorescence we should observe experimentally. In Figure 3a, which corresponds to a Z dipole, the curves for $60° \leq \theta \leq 90°$ best resemble the representative plot of Figure 2d. This range of $\theta$ also corresponds to the brightest defect orientations and is therefore the most likely set of orientations to identify experimentally. In Figure 3b, which corresponds to an XY dipole, the plots for $60° \leq \theta \leq 90°$ are also the most similar to the experimental plot of Figure 2d. However, for an XY dipole, this range of $\theta$ corresponds to the dimmest defect orientations and is therefore the least likely set of orientations to be encountered experimentally. Thus in our Z (XY) dipole model, only the defects we are most (least) likely to encounter experimentally match our observations. Analogous $\theta$-dependent plots for the remaining measurements of Figures 2a, b, and c similarly support a Z dipole over an XY dipole.

As a final verification that we study a Z dipole, we examine the distribution of polarization visibilities. The visibility is defined as $V = (I_{max} - I_{min})/(I_{max} + I_{min})$, where $I_{max}$ and $I_{min}$ are the maximum and minimum fluorescence intensities. The distribution of excitation and emission visibilities from defects in randomly oriented NPs is shown in Figures 4a and b, respectively. The excitation (emission) visibilities range from 0.50 (0.49) to 0.91 (0.94) and have an average value of 0.78 (0.79). Superposed on the visibility distributions is a best-fit produced from our Z (solid line) and XY (dashed line) dipole models. These best-fits employ the same binning width as the data and incorporate a background parameter that left-shifts the theoretical distribution. The Z dipole model better fits the data than the XY model and also predicts a background that better matches that seen experimentally.

Figures 2-4 establish that the investigated defects possess a Z dipole. Typically the associated absorption and emission dipoles for point-defects are nearly parallel.[8,39,41,42] An earlier report[18] found a large (~80°) offset between the absorption and emission maxima of a quantum emitter in ZnO. Figure 4c shows our distribution of misalignment between the absorption and emission maxima for all defects investigated. The distribution is concentrated near 0° and the average misalignment is 3.2°. We interpret this small misalignment as a systematic error in the average polarization of the collection path that results from variations in path retardance for collected photons whose wavelength differs from our 630 nm calibration wavelength. Consequently we find that the absorption and emission dipoles are parallel within experimental uncertainty.

Once a defect's dipole type (Z or XY) is known, group theoretic considerations can shed light on the ground and excited state wavefunction properties.[43] Every point-defect belongs to a point-group containing the symmetry operations that leave its Hamiltonian invariant. For the ZnO wurtzite lattice, the available defect symmetry operations are the identity ($\hat{E}$), a 120° rotation about the c-axis ($\hat{C}_3$), and a reflection about a vertical plane ($\hat{\sigma}_v$). These operations yield three nontrivial point-groups: $\boldsymbol{C_{3v}} = \{\hat{C}_3, \hat{\sigma}_v, \hat{E}\}$, $\boldsymbol{C_s} = \{\hat{\sigma}_v, \hat{E}\}$, and $\boldsymbol{C_3} = \{\hat{C}_3, \hat{E}\}$. Table 1 lists the allowed transitions for these point-groups for each dipole type (see Supplementary Information). If the point-group of the defects we study is among these, then both the ground and excited state wavefunctions must be orbital singlets and transform as the $A_1$ or $A_2$ irreducible representation. Previous observations of photon bunching[16–18] revealed that SPSs in ZnO possess a metastable shelving state. Of the point-groups in Table 1, only $\boldsymbol{C_{3v}}$ has a forbidden transition ($A_1 \leftrightarrow A_2$) that becomes a candidate for this metastable to ground state transition. Note that an orbital singlet may possess spin degrees of freedom, as is the case for the orbital singlet, spin triplet ground state of the NV center.

We note that our study does not rule out the unlikely possibility that the investigated defects belong to a point-group not listed in Table 1. In such a scenario the defect would not stem solely from an impurity, vacancy, or a small combination of these. While a Z dipole is never of the $A \leftrightarrow E$ variety, it can correspond to $E \leftrightarrow E$ transitions in some high symmetry cases such as the tetragonal point-group $D_{4h}$. That said, these cases are unlikely to be relevant in the bulk or surface of ZnO and the most likely scenario is a transition between orbital singlets. Future measurements in single crystal ZnO can test these possibilities to aid in identifying the crystallographic direction of the defect's symmetry axis.

In conclusion, we investigated polarization properties of defect-based SPSs in ZnO to gain insight into their electronic states and their structural origin by discerning whether they possess a Z or XY dipole. Because particular orientations of an XY dipole resemble a Z dipole, and vice versa, experiments intended to identify a SPS's dipole type have previously been conducted in ordered crystals where the allowed orientations of the emitter are known.[8,41,44,45] Here we demonstrate that an alternative but definitive approach is to sample emitters that are randomly oriented and compare the distribution to the statistical expectation for each dipole type. Though we apply this approach to point-defects in randomly oriented nanoparticles, it could also be useful for establishing the dipole type of unknown quantum emitters like single molecules in amorphous media where the orientation is not *a priori* known.  For each single defect investigated in this work, either a Z or XY dipole model could replicate the observed behavior.  However, the XY model fails at predicting the distribution of observed polarization plots and visibilities, while the Z model satisfactorily matched these distributions.  Consequently, our measurements and analysis indicate that the investigated defects have a Z dipole.

From polarization measurements of 32 distinct defects, we find that the absorption and emission dipoles are parallel within experimental uncertainty. This indicates that the symmetry axis of the ground state is identical to that of the excited state within the precision of our measurement.  Though the defect's absolute symmetry remains unknown, we have applied group theoretic considerations to all crystallographic point-groups.  We find that if the defect belongs to a point-group relevant to ZnO then the ground and excited states are orbital singlets.  Establishing that the investigated ZnO defects possess a Z dipole and identifying their most probable electronic states constitutes an important step in pinpointing their structural origin.

**Acknowledgements:**

This work was supported by the National Science Foundation (DMR-1254530). We acknowledge use of the Cornell NanoScale Facility, a member of the National Nanotechnology Coordinated Infrastructure (NNCI), which is supported by the National Science Foundation (Grant ECCS-15420819).  Additionally, we acknowledge the Cornell Center for Materials Research Shared Facilities, which are supported through the NSF MRSEC program (DMR-1120296).

**Table 1:**

|  | $C_{3v} = \{\hat{C}_3, \hat{\sigma}_v, \hat{E}\}$ | $C_s = \{\hat{\sigma}_v, \hat{E}\}$ | $C_3 = \{\hat{C}_3, \hat{E}\}$ |
|---|---|---|---|
| Z Dipole | $A_1 \leftrightarrow A_1$, $A_2 \leftrightarrow A_2$ | $A_1 \leftrightarrow A_2$ | $A_1 \leftrightarrow A_1$ |
| XY Dipole | $A_1 \leftrightarrow E$, $A_2 \leftrightarrow E$ | $A_1 \leftrightarrow A_1$, $A_2 \leftrightarrow A_2$ | $A_1 \leftrightarrow E$ |

Summary of allowed transitions for Z and XY dipoles for point-groups relevant to ZnO. States labeled by "$A$" ("$E$") are singlets (doublets). The "1" subscript indicates the state is totally symmetric and the "2" subscript indicates the state is antisymmetric with respect to $\hat{\sigma}_v$. Because transitions are symmetric, the ground and excited states in a given transition may be interchanged.

**Figure 1:**

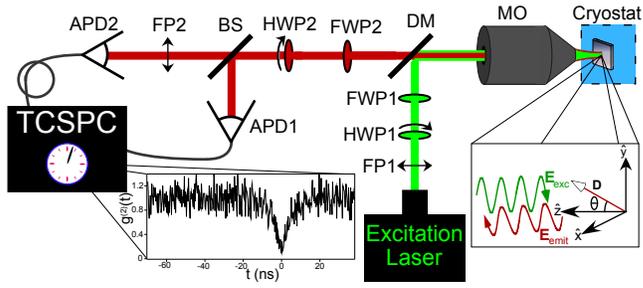

Schematic of confocal microscope used to investigate polarization properties of single defects in ZnO.

**Figure 2:**

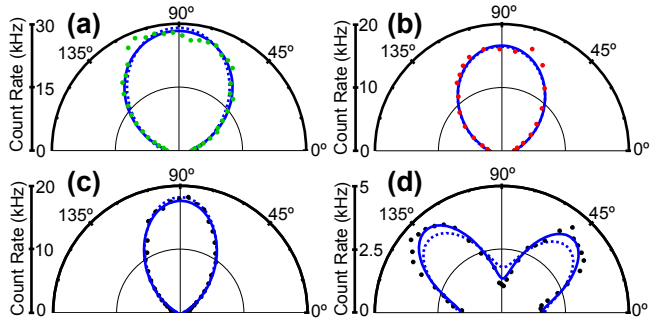

Representative set of polarization measurements acquired for single defects in this work. The absorption (a) and emission (b) graphs are aligned, suggesting the absorption and emission dipoles are parallel. The smaller count rate in (b) compared with (a) results from differences in collection efficiency for the two paths. Plots (c) and (d) corresponds to the polarization of the exciting light and the collected light being parallel and perpendicular, respectively. The solid (dashed) line is the best-fit produced assuming a Z (XY) dipole. The polarization basis in these plots has been defined so that the excitation maximum points along the y-axis.

**Figure 3:**

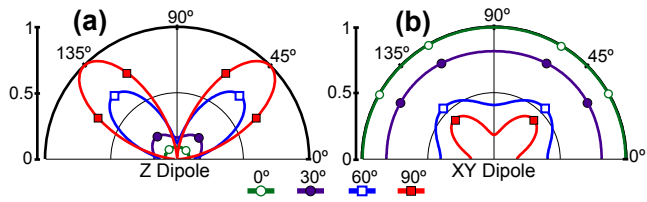

Theoretical plots for the case when the polarization of the exciting light and collected light are perpendicular for a Z dipole (a) and an XY dipole (b). Each graph has four curves that correspond to various symmetry axis orientations.

**Figure 4:**

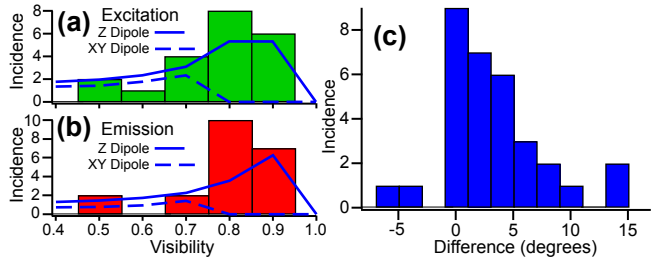

Histogram of the excitation visibility (a), emission visibility (b), and angular difference between the absorption and emission maxima (c). Superposed in (a) and (b) are best-fits resulting from our Z (solid) and XY (dashed) dipole models.

# Supplemental Material: "Polarization spectroscopy of defect-based single photon sources in ZnO"

N. R. Jungwirth, H. -S. Chang, M. Jiang, and G. D. Fuchs

Cornell University, Ithaca, New York 14853, USA

## I. Sample Details

The samples consist of randomly oriented nanoparticle (NP) ZnO and ⟨001⟩ oriented sputtered ZnO films. The NPs were suspended in methanol prior to being drop-cast onto a thermally oxidized silicon substrate and are commercially available from Sigma-Aldrich as product number 677450. The sputtered films were also deposited on thermally oxidized silicon and the growth parameters have been described previously.[1] All samples were annealed in air at $500°\,C$ for 30 minutes and then passivated with the e-beam resist hydrogen silsesquioxane (HSQ). The HSQ layer was motivated by a previous report[2] and phenomenologically serves to increase the photo-stability of SPSs. A control sample containing HSQ, but not ZnO, verified that the investigated SPSs originate from ZnO. All data was obtained for samples mounted in a Janis ST-500 cryostat capable of cooling to 4.5 K. We detect no variation in polarization properties from 4.5 K to room temperature.

## II. Excitation using 473nm light

The absorption band of isolated defects is often broadened by phonon coupling. In addition to using 532 nm (green) light, we also investigated using 473 nm (blue) light for excitation. We detected no difference in defect polarization or emission properties when using blue light in lieu of green light. We did, however, observe an increase in the background fluorescence when using blue light compared to green light.

Figure 1 shows $g^{(2)}(\tau)$ for a ZnO defect measured using 532 nm and 473 nm light for excitation. In each plot, $g^{(2)}(\tau = 0) < 0.5$, indicating that a single defect is being probed. The antibunching dip for 473 nm excitation is not as deep as the dip for 532 nm light because of the increase in the uncorrelated, background fluorescence relative to defect fluorescence. Additionally, both curves exhibit photon bunching $[g^{(2)}(|\tau| \sim 10\ ns) > g^{(2)}(\infty)]$, establishing that the defect has a metastable shelving state.

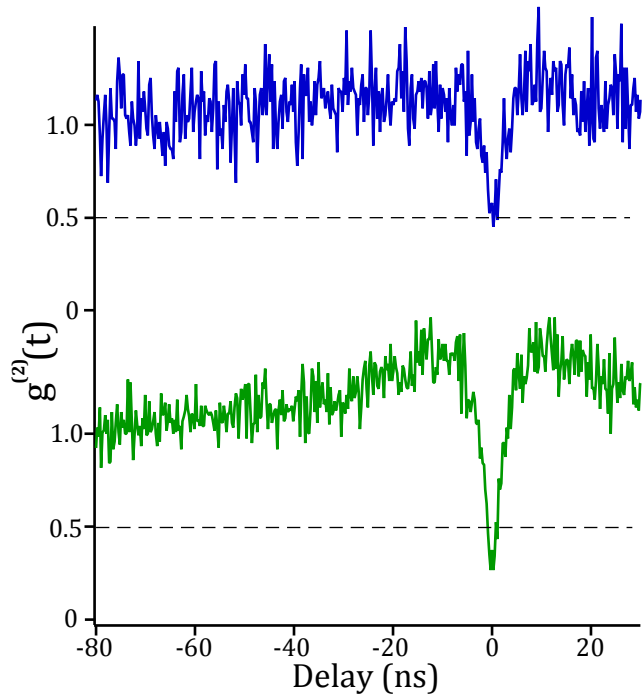

**Figure 1:** Measurement of $g^{(2)}(\tau)$ using blue (top) and green (bottom) light for excitation

Figure 2 compares the normalized emission spectrum of a single defect in ZnO excited with blue and green light. The two spectra are nearly indistinguishable, suggesting that both blue and green light excite the defect into the same excited electronic state.

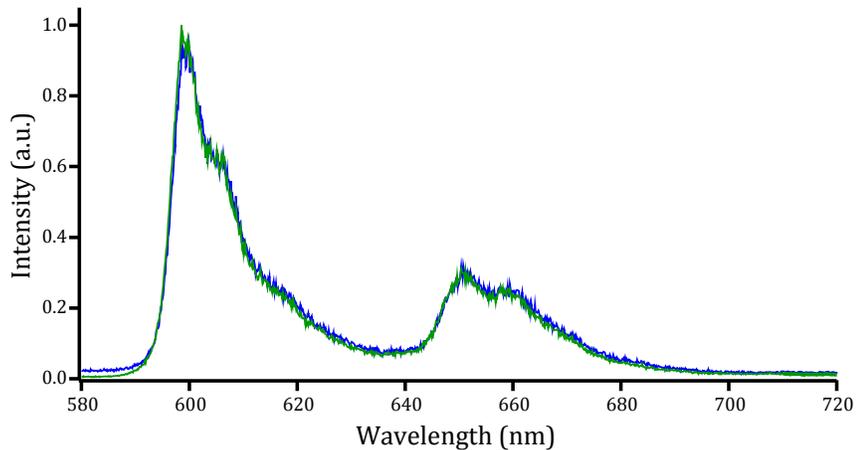

**Figure 2:** Normalized emission spectrum of a single ZnO defect excited using blue and green light. The emission spectrum is not altered by the wavelength of the exciting light.

Figure 3 displays a set of single defect polarization measurements made using blue and green light for excitation. The polarization basis has been defined so that the maximum of the green absorption curve (green triangles) occurs at 90°. The maximum of the blue absorption curve (blue diamonds) is also at ~90°, indicating that the absorption dipoles for both blue and green light are identical. Lastly, the defect emission polarization (red circles) also presents a maximum at ~90° because the absorption and emission dipoles are parallel.

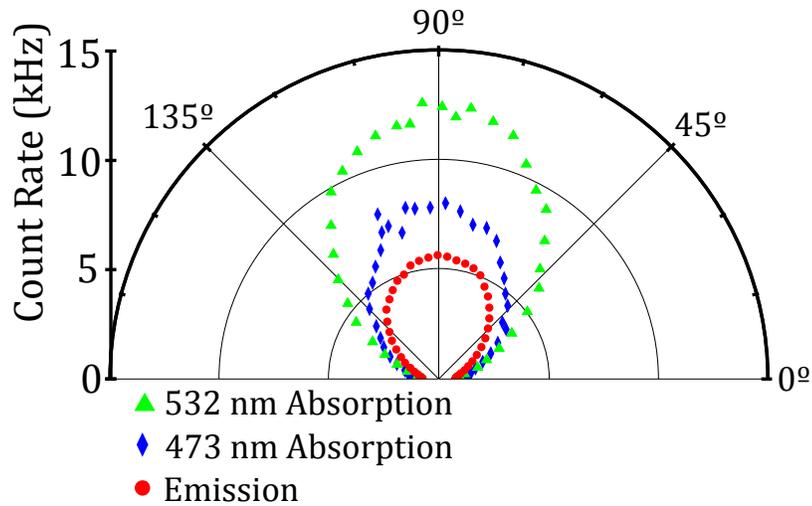

**Figure 3:** Set of polarization measurements made on a single defect using blue and green light for excitation.

### III. Temperature Dependence

We measured the orientation of a single defect's absorption dipole at several temperatures between 10 and 90 K. Figure 4 shows the result of this experiment. No temperature dependence is evident, and the maximum variation is 1.7°, which is within our experimental uncertainty, suggesting the defect's dipole orientation remains stable as temperature is varied.

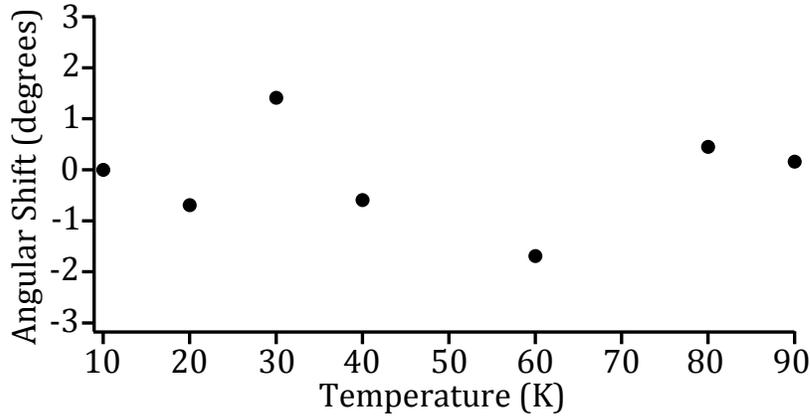

**Figure 4:** Angular shift of the absorption dipole of a single defect from 10 to 90 K. No temperature dependence is evident.

## IV. Point-group theory

If a defect's point-group is known then valuable information about its eigenstates and polarization selection rules may be extracted without computation. A thorough overview of group theory and its applicability to physics is available in Dresselhaus.[3] Here we quickly illustrate how we applied point-group theory to our findings by considering the point-group $C_{3v}$.

Table 1 is the $C_{3v}$ character table.[3] The leftmost column of Table 1 lists the irreducible representations ($A_1$, $A_2$, and $E$) associated with $C_{3v}$. There is a one-to-one correspondence between the irreducible representations of a defect's point-group and the single electron eigenstates of its Hamiltonian. The dimensionality of the representation corresponds to the eigenstate's degeneracy and the eigenstate possesses the transformation properties of its corresponding irreducible representation. The $A_1$ ($A_2$) representation is one-dimensional and is symmetric (antisymmetric) under the three $\hat{\sigma}_v$ reflections, whereas the $E$ representation is two-dimensional. Consequently, defects with $C_{3v}$ symmetry have two non-degenerate states and one doubly degenerate state. Spin properties arise when these single electron states are populated with electrons in accordance with the Pauli exclusion principle.

The last column of Table 1 indicates how the linear operators x, y, and z transform. Thus for $C_{3v}$, the z operator transforms as the totally symmetric representation $A_1$ whereas x and y jointly transform as the two-dimensional $E$ representation. Note that, contrary to our coordinate system used previously, the z direction here is aligned parallel to the defect's symmetry axis rather than parallel the axis of the microscope objective.

|       | $\hat{E}$ | $2\hat{C}_3$ | $3\hat{\sigma}_v$ | Linear |
|-------|-----------|--------------|-------------------|--------|
| $A_1$ | 1         | 1            | 1                 | z      |
| $A_2$ | 1         | 1            | -1                |        |
| $E$   | 2         | -1           | 0                 | x,y    |

**Table 1:** Character table for the point-group $C_{3v}$.

In general, a transition between initial state $|\psi_i\rangle$ and final state $|\psi_f\rangle$ is allowed if $\langle\psi_f|\hat{O}|\psi_i\rangle \neq 0$, where $\hat{O}$ is a perturbation intended to couple $|\psi_i\rangle$ and $|\psi_f\rangle$. In the case of an electric dipole transition from the ground to excited state, $\hat{O} = \boldsymbol{\mu} \cdot \boldsymbol{E}$, where $\boldsymbol{\mu}$ is the absorption dipole and $\boldsymbol{E}$ is the electric field of the exciting light. Consequently, in the basis of the defect, light polarized along $(x, y, z)$ may be absorbed and excite the defect if $\langle\psi_f|(x,y,z)|\psi_i\rangle \neq 0$. Group theory may be utilized to quickly determine which of these transition probabilities necessarily vanish. If the ground state $|\psi_i\rangle$, excited state $|\psi_f\rangle$, and perturbation $\hat{O}$ transform as the irreducible representations $\Gamma_i$, $\Gamma_f$, and $\Gamma_O$, respectively, then a transition is permitted by group theory if and only if the direct product $\Gamma_i \otimes \Gamma_o \otimes \Gamma_f$ contains the totally symmetric representation.

Table 2 is a direct product table for $C_{3v}$ and is useful for determining polarization selection rules of a point-defect with $C_{3v}$ symmetry. For example, z-polarized light may drive $A_1 \leftrightarrow A_1$ transitions because $A_1 \otimes \Gamma_z \otimes A_1 = A_1 \otimes A_1 \otimes A_1 = A_1 \supset A_1$. However, z-polarized light cannot drive $A_1 \leftrightarrow A_2$ transitions because $A_1 \otimes A_1 \otimes A_2 = A_2 \not\supset A_1$. Table 3 lists the transitions permitted by group theory for x, y, and z polarized light and can be used to determine the allowed transitions for a Z and XY dipole. To illustrate, a Z dipole is associated with $A_1 \leftrightarrow A_1$ and $A_2 \leftrightarrow A_2$ transitions because only *one* polarization state drives these transitions. Moreover, an XY dipole is associated with $A_1 \leftrightarrow E$ and $A_2 \leftrightarrow E$ transitions because precisely *two* orthogonal polarization states drive these transitions. Transitions of the $E \leftrightarrow E$ variety are permitted for x, y, and z polarized light and would therefore be associated with an XYZ dipole.

| $\otimes$ | $A_1$ | $A_2$ | $E$ |
|---|---|---|---|
| $A_1$ | $A_1$ | $A_2$ | $E$ |
| $A_2$ | $A_2$ | $A_1$ | $E$ |
| $E$ | $E$ | $E$ | $A_1 + A_2 + E$ |

**Table 2:** Product table for the irreducible representations associated with the point-group $C_{3v}$.

| Polarization | Allowed Transitions |
|---|---|
| x ,y | $A_1 \leftrightarrow E, A_2 \leftrightarrow E, E \leftrightarrow E$ |
| z | $A_1 \leftrightarrow A_1, A_2 \leftrightarrow A_2, E \leftrightarrow E$ |

**Table 3:** List of transitions that are permitted by group theory for light polarized along the x, y, and z directions.